\documentclass[useAMS,usenatbib]{mn2e}
\usepackage{graphicx, setspace, subfigure, latexsym, amssymb, amsmath, booktabs, wasysym, paralist}
\usepackage{multirow}
\usepackage{eqnarray}
\usepackage{gensymb}

\title[PsrPopPy]{PsrPopPy: An open-source package for pulsar population simulations}
\author[S. D. Bates et al.]{S. D. Bates$^{1,2}$, D. R. Lorimer$^{1,3}$, A. Rane$^{1}$ and J. Swiggum$^{1}$\\
$^{1}$Department of Physics and Astronomy, West Virginia University, Morgantown, WV, 26506 USA\\
$^{2}$Jodrell Bank Centre for Astrophysics, School of Physics and Astronomy, The University of Manchester, Manchester M13 9PL, UK\\
$^{3}$National Radio Astronomy Observatory, PO Box 2, Green Bank, WV 24944, USA
}
\begin{document}

\date{Accepted; \today}

\pagerange{\pageref{firstpage}--\pageref{lastpage}} \pubyear{2012}

\maketitle

\label{firstpage}
\begin{abstract}
We have produced a new software package for the simulation of pulsar populations,
\textsc{PsrPopPy}, based on the \textsc{Psrpop} package.
The codebase has been re-written in Python (save for some external libraries, which 
remain in their native Fortran), utilising the object-oriented features of the language,
and improving the modularity of the code. Pre-written scripts are provided
for running the simulations in `standard' modes of operation, but the code 
is flexible enough to support the writing of personalised scripts. The modular
structure also makes the addition of experimental features (such as new models for
period or luminosity distributions) more straightforward than with the previous code.
 We also discuss potential
additions to the modelling capabilities of the software. Finally, we demonstrate some
potential applications of the code; first, using results of surveys at different observing
frequencies, we find pulsar spectral indices are best fit by a normal distribution with
mean $-1.4$ and standard deviation $1.0$. Second, we model pulsar spin evolution to 
calculate the best-fit for a relationship between a pulsar's luminosity and
spin parameters. We used the code to replicate the analysis of Faucher-Gigu\`ere \& Kaspi,
and have subsequently optimized their power-law dependence of radio luminosity, $L$,
with period, $P$, and period derivative, $\dot{P}$. We find that the underlying
population is best described by $L \propto P^{-1.39 \pm 0.09} \dot{P}^{0.48 \pm 0.04}$
and is very similar to that found for $\gamma$-ray pulsars by Perera et al. 
Using this relationship, we generate a model population and examine
the age-luminosity relation for the entire pulsar population, which may be 
measurable after future large-scale surveys with the Square Kilometer Array.

\end{abstract}


\section{Introduction}\label{sec:intro}
The known Galactic population of pulsars stands at over 2000 sources, but represents
only a small fraction of the total detectable Galactic
population of $\sim 120 000$ \citep[][hereafter FK06]{fk06}. 
In order to understand and make predictions about the unseen population, simulations
are required to disentangle the large number of competing effects.
One technique for simulating the Galactic population is to take the 
observed population and, providing the biases in that dataset are understood correctly, 
extrapolate to the whole Galaxy. This is known as the `snapshot' method, since rather
than evolving the pulsars from some initial conditions, it simply provides a 
representation of the population as it stands at this point in time.

\citet{lfl+06} used results from the most successful pulsar survey to date, the 
Parkes Multibeam Pulsar Survey \citep[][]{mlc+01}, and the high-latitude pulsar survey
\citep[][which used the same observing system]{bjd+06} to form a dataset comprising of
1008 pulsars. Using these pulsars, they were able to form models of period, luminosity
and spatial distributions for pulsars in the Galaxy. This software used in this work 
has since been made available to the pulsar community as 
\textsc{Psrpop}\footnote{http://psrpop.sourceforge.net/}, and as one of the few
publicly-available population simulation software packages has been commonly used
as a tool either for predicting the results of future pulsar surveys \citep[e.g.\,][]{keith2010} or for study of some feature of the pulsar population
\citep[e.g.\,][]{ridley2010, blv13}.

In this paper we present a new version of this software, rewritten in the Python
programming language, providing increased modularity, accessibility for new 
contributors, and making it easier to add new features than in the Fortran codebase. 
In \S~2 we outline the various formulae used to create models
in the snapshot code. 
\S~3 discusses the formulae used to generate simulated populations in the
evolutionary code, and in \S~4 we summarise the method for generating pulsar population
models using these formulae. \S~5 uses the code to address some sample problems in 
population statistics, and finally we draw conclusions and suggest some further
work to be done.

\section{The Snapshot Method}\label{sec:snapshot}
\textsc{PsrPopPy} uses a series of statistical models in order to generate a 
simulated Galactic pulsar population. These models describe the pulse period, 
luminosity, and spatial distributions, and the currently supported models are 
described below. Since the pulsar population has been so 
well studied at radio frequencies of 
$\sim 1.4~\mathrm{GHz}$, all frequency-dependent parameters are initialised 
using this reference frequency, and then extended to other radio frequencies 
through the use of power law spectral indices.

\subsection{Pulse Period Distributions}
\subsubsection*{Normal distribution}
A logical starting point in many studies may be to use a normal distribution
to describe pulsar periods. Any values can be given for the mean and standard
deviation of this distribution, and periods are drawn from it randomly. FK06 
used a normal distribution to describe the birth spin periods of pulsars in their 
simulations of pulsar evolution.

\subsubsection*{Log-normal distribution}
A more popular model for the current distribution of pulsar spin periods is to
use the log-normal distribution. \citet{lfl+06} found the best-fit of this model
to the known population of non-recycled pulsars 
has mean $\mu = 2.7$, and standard deviation
$\sigma = -0.34$ (in logarithm to the base-10), used as default values in \textsc{PsrPopPy}. The end point of the 
simulations mentioned above, by FK06, was also a log-normal period distribution;
that is, the value of $x \equiv \log_{10}P$ is drawn from the distribution,

\begin{equation}
f(x;~\mu,\sigma) = \frac{1}{\sqrt{2\pi\sigma^2}}e^{-(x - \mu)^2/(2\sigma^2)}.
\label{eq:lognorm}
\end{equation}

\subsubsection*{Millisecond pulsars}
Although outside the `normal usage' of \textsc{PsrPopPy}, the simulation of 
Galactic millisecond pulsars (MSPs) is an important problem to address, especially 
as many modern pulsar surveys are aimed at the discovery of MSPs with the 
potential for extremely high timing precision \citep[see, e.g.\,][]{keith2010, batesmsps,
boyles2012}.

Currently, \textsc{PsrPopPy} provides the MSP period model devised by \citet{cc97} 
which can be used to model MSP periods. However, \citet{lorimerIAU2012} suggests that
this period distribution does not well describe the currently known population of
MSPs, which has grown substantially since that model was proposed. We intend to 
add more MSP period models based upon future results.

\subsection{Modelling pulse widths}
\subsubsection*{Log-normal distribution}
The most simple pulse width model draws pulse widths from a log-normal 
distribution with standard deviation 0.3 in pulse phase, and a mean given by the user.

\subsubsection*{Using a model of pulsar beam widths}
The following discussion describes the pulsar beam model first described for use in
pulsar simulations by \citet{smits2009}, which assumes a simple geometry --- pulsars
in this model have circular beams of radius $\rho$, the angle between the pulsar's 
spin and magnetic axes is given by $\chi$ and the angle between the magnetic axis and
the line of sight to Earth is $\beta$. \citet{kxc+98} found the following empirical 
relation between the value of $\rho$ and spin period $P$;

\begin{equation}
    \rho = 
        \begin{cases}
            5.4\degree \text{ } P^{-1/2} & \text{, } P>30\mathrm{ms} \\
            31.2\degree & \text{, } P\leq 30\mathrm{ms}\\
        \end{cases}
    \label{eq:kramer98}
\end{equation}
which is used to generate initial values of $\rho_\mathrm{init}$. 
The values are then dithered by a value, 
$p$, drawn from a uniform distribution between $-0.15$ and $0.15$,

\begin{equation}
    \log_{10} \rho = \log_{10} \rho_\mathrm{init} + p.
    \label{eq:ditherrho}
\end{equation}

The value of $\beta$ is then chosen from a uniform distribution as 
$-\rho \leq \beta \leq \rho$, and $\chi$ is calculated by

\begin{equation}
    \chi = \arccos q,
    \label{eq:chi}
\end{equation}
where $q$ is selected, again, from a uniform distribution such that $0<q<1$. Using
these values of $\chi$, $\beta$ and $\rho$,
the equation

\begin{equation}
    \sin^2\left(\frac{W}{4}\right) = \frac{\sin^2 \left(\frac{\rho}{2}\right) - 
    \sin^2 \left(\frac{\beta}{2}\right)}
                                {\sin \alpha \sin (\alpha+\beta)}
    \label{eq:pulsewidth}
\end{equation}
is used to calculate the pulse width \citep[e.g.\,][]{gil1984}.

\subsection{Luminosity Distributions}
\subsubsection*{Power law}
Use of a power-law to describe the distribution of pulsar luminosities has been 
common since studies of the pulsar population began. For example, \citet{tm77} fitted a 
power-law with index $\mathrm{d} \log_{10} N / \mathrm{d} \log_{10} L =-1.12(5)$. 
The study by \citet{lfl+06} also used a simple power-law when describing
the luminosity distribution, obtaining a slightly shallower power-law index
of $\sim -0.7$ (depending upon the model chosen). However, the use of a power-law
requires a cut-off luminosity to prevent extremely high luminosities; this
is unsatisfactory from a consideration of the physics involved, and has led to 
the adoption of log-normal distributions in more recent work.

\subsubsection*{Log-normal distribution}
To avoid using a luminosity cut-off \citep[as discussed by][]{ridley2010},
pulsar luminosities are drawn from a log-normal distribution (see 
Equation~\ref{eq:lognorm}) with a mean of $\langle \log_{10} L \rangle = -1.1$ 
and standard deviation
$\sigma_{\log_{10} L} = 0.9$, following FK06.

\subsubsection*{Parameterised in $P$ and $\dot{P}$}\label{sec:lumlaw}
For evolutionary models where a pulsar is given both a rotation period, $P$, and
a period derivative, $\dot{P}$, the radio luminosity may be parameterised as
see, e.g.\,FK06,

\begin{equation}
    \log_{10} L = \log_{10} ( P^\alpha \dot{P}_{15}^\beta \gamma) + L_\mathrm{corr},
\label{eq:ppdotlum}
\end{equation}
where $L_\mathrm{corr}$ is a dithering factor chosen from a normal distribution
centred on zero with a variable standard deviation.

\subsection{Spatial Distribution}
\subsubsection*{Simple models}
Three basic models are provided for distributing the pulsars around the Galaxy.
In the \textbf{disk} model, pulsars are distributed along the 
Galactic plane ($z=0$), with random Galactic $x$-$y$ coordinates from $-15$ to
$+15$~kpc. In the \textbf{slab} model, Galactic $x$ and $y$ coordinates 
are calculated as with the disk distribution, but $z$-coordinates run 
from $-5$~kpc to $+5$~kpc. Finally, in a very simple \textbf{isotropic} model, pulsars 
are distributed randomly around the Earth at a distance of 1~kpc. 

\subsubsection*{Radial distribution}
More complex radial distribution models which attempt to take into account the structure 
of the Galactic disk have also been incorporated into \textsc{PsrPopPy}; the first is 
the radial model produced by \citet{yk2004}. The second, and default, is an adaptation 
of this model based upon the analysis of \citet{lfl+06}. Thirdly, pulsars can be distributed around the Galactic centre using a Gaussian radial density profile of 
variable width \citep[e.g.\,][]{narayan1987}.

\subsubsection*{Galactic scale height}\label{sec:scaleheight}
The distribution of pulsars in Galactic $z$ coordinates is commonly approximated 
by a two-sided exponential \citep[e.g.\,][]{lyne1998, lfl+06}. The default value
used in \textsc{PsrPopPy} is 330~pc, as obtained by \citet{lfl+06}, but can be
set at any value --- for instance, the scale height for MSPs is considerably higher 
than for non-recycled pulsars. Recent work by \citet{levin2013} obtains a best-fit
MSP scale height of 500~pc.
\textsc{PsrPopPy} also supports the use of Gaussian 
distributions with mean of zero and variable width to model pulsar Galactic scale height.

\subsection{Galactic Electron Distribution}\label{sec:electrondist}
Two popular models for the conversion between dispersion measure (DM; the column density 
of free electrons along a given line of sight) and true distance are included in
\textsc{PsrPopPy}. The popular \citet{ne2001} model is included which is, to date, 
the best available model for obtaining distance estimates for pulsars at a given DM. 
The library for performing this transformation has been kept in
Fortran; sometimes it can be prohibitively slow, so the model
from \citet{lmt85} is also included. The recent models by \citet{schnit2012} are not yet
included, but work to include these and other electron density models is ongoing.

\subsection{Modelling Scintillation Effects}\label{sec:scintillation}
We model variations in pulse intensity, known as scintillation, caused by
the relative motions of the Earth, pulsars and the interstellar medium. Our discussion
of this effect is based upon the outline given by \citet{lk05}. 

If a pulsar 
of flux $S$ is seen to vary with standard deviation $\sigma_\mathrm{s}$, the
intensity of these variations is described by the modulation index, $m$,
\begin{equation}
    m = \sigma_\mathrm{s} / \langle S \rangle,
\label{eq:modindex}
\end{equation} 
whose value must be computed at observing frequency $f$ from the 
scintillation strength, 
\begin{equation}
    u = \sqrt{\frac{f}{\Delta f_\mathrm{DISS}}},
\label{eq:scintstrength}
\end{equation}
where $\Delta f_\mathrm{DISS}$ is the diffractive
scintillation bandwidth, given by
\begin{equation}
    \Delta f_\mathrm{DISS} = \frac{2 \pi \tau_\mathrm{s}}{C_1},
\label{eq:scintbw}
\end{equation}
where $\tau_\mathrm{s}$ is the scattering time, given by Equation~\ref{bhatscatter}, 
and we assume a Kolmogorov spectrum, and hence the constant $C_1 = 1.16$.

In the regime where $u<1$, known as weak scintillation, the modulation index
is simply computed as
\begin{equation}
    m = \sqrt{u^{5/3}}, 
\label{eq:weakscint}
\end{equation}
but for $u>1$, strong scintillation, the situation becomes more 
complicated, and we must compute modulation indices for both refractive and 
diffractive scintillation, $m_\mathrm{RISS}$ and $m_\mathrm{DISS}$ respectively. The 
total modulation index is then computed as 
\begin{equation}
    m^2 = m_\mathrm{RISS}^2 + m_\mathrm{DISS}^2 + m_\mathrm{RISS}m_\mathrm{DISS}.
\label{eq:strongscint}
\end{equation}

The modulation index of refractive scintillation is simply related to the
scintillation strength as 
\begin{equation}
    m_\mathrm{RISS} = u^{-1/3}, 
\label{eq:mriss}
\end{equation}
while the modulation index for diffractive scintillation is related to the number
of scintles, $N$, sampled in the observing time $\Delta t_\mathrm{obs}$ 
and observing bandwidth $\Delta f_\mathrm{obs}$,
\begin{equation}
    m_\mathrm{DISS} = \frac{1}{\sqrt{N_\mathrm{t} N_\mathrm{f}}}, 
\label{eq:mdiss}
\end{equation}
where
\begin{equation}
    N_\mathrm{t} \approx 1 + \kappa\frac{\Delta t_\mathrm{obs}}{\Delta t_\mathrm{DISS}}
\label{eq:N_t}
\end{equation}
and
\begin{equation}
    N_\mathrm{f} \approx 1 + \kappa\frac{\Delta f_\mathrm{obs}}{\Delta f_\mathrm{DISS}}
\label{eq:N_f}
\end{equation}
and we take an average value of $\kappa = 0.15$. In the strong regime, we rely on 
the NE2001 model to compute the values of $\Delta f_\mathrm{DISS}$ and 
$\Delta t_\mathrm{DISS}$ \citep{ne2001}.

In the cases of both weak and strong scintillation, the value of $m$ is then
applied to the pulsar (for which flux $S$ has already been calculated based upon
the distance to the pulsar and its luminosity). Following the definition of 
the modulation index in Equation~\ref{eq:modindex}, the value of $\sigma_\mathrm{s}$
can be computed. A modified flux value is then chosen at random from a normal 
distribution with mean $S$ and standard deviation $\sigma_\mathrm{s}$, and is applied
only for the current realisation in the simulation. The original value of $S$ is stored
so that the modulation will vary each time code is used to simulate a pulsar
survey (see \S~\ref{sec:dosurvey}).

\subsection{Spectral Index Distribution}\label{sec:specindexintro}
\textsc{PsrPopPy} allows radio spectral indices to be normally distributed with a given
mean and standard deviation $\alpha$ and $\beta$, respectively. 
The default values are $\alpha = -1.6,~\sigma = 0.35$
\citep{lorimer1995}, however, a study by \citet{maron2000} derived $\alpha = -1.8
, \sigma=  0.2$, and further work by \citet{blv13} finds
an underlying spectral index distribution given by $\alpha = -1.4,~\sigma = 1.0$.

\section{Modelling pulsar spin evolution}\label{sec:evolve_algorithms}
For evolutionary simulations of the pulsar population, \textsc{PsrPopPy} includes 
models for generating period derivatives for each pulsar, based upon work by 
FK06 and \citet{cs06}. The method follows that discussed by \citet{ridley2010}
and includes a variety of pulsar beaming models and the option of including a decay in 
the angle between the spin and magnetic axes of each pulsar. 

Although the evolution code is contained in a separate executable, the output models
are simply serialised versions of the models stored in memory. This format is 
identical to that used in the rest of \textsc{PsrPopPy}, and so the models are completely
compatible with the output from the snapshot simulations. As well as the distributions
outlined in \S~\ref{sec:snapshot}, additional parameters
need to be modelled, using the distributions described below.

\subsection{Magnetic Field Distribution}
By default, the pulsar magnetic fields are selected from a log-normal distribution 
with mean $\langle \log_{10} B \rangle = 12.65$ and standard deviation
$\sigma_{\log_{10} B} = 0.55$. These values are chosen per FK06, but may be altered.

\subsection{Rotational Alignment Distributions}
Models of the pulsar spindown, discussed in \S~\ref{subsec:spindownmodels}, make use
of the angle, $\chi$, between the rotational and magnetic axes of the pulsar. 

The most simple model treats all pulsars as orthogonal rotators; that is, $\chi=90\degree$
for every simulated pulsar, while a more realistic model aligns the spin 
and magnetic axes at random. The value of $\chi$ 
is calculated in the same way as in Equation~\ref{eq:chi}.


Additionally, a beaming model is implemented based upon the the work of 
\citet{wj08} who found evidence for an alignment of the magnetic and rotational axes
over a timescale $t_\mathrm{d} \sim 7 \times 10^7\mathrm{~yr}$. Therefore, we include
the possibility of $\chi$ decaying from an initial value $\chi_0$ as 

\begin{equation}
    \sin \chi = \sin \chi_0 \exp(-t/t_\mathrm{d})
\label{eq:wj08}
\end{equation}
after a time $t$. Following \citet{ridley2010}, this formula does not exactly
replicate the model as described by \citeauthor{wj08}, but describes a simplified
version.

\subsection{Modelling pulsar spindown}\label{subsec:spindownmodels}
\subsubsection*{Magnetic dipole model}
The general expression for the pulse period, $P$ as a function of time in the magnetic
dipole model \citep[see][for details]{ridley2010} for a pulsar with braking index $n$
is given by 
\begin{equation}
    P(t) = \left[ P_0^{n-1} + \frac{n-1}{2} t_\mathrm{d}kB^2 
                    \sin^2 \chi_0 (1 - \exp(-2t/t_\mathrm{d}))\right]^\frac{1}{n-1} 
\label{eq:p_t_ridley}
\end{equation}
where the constant

\begin{equation}
    k = \frac{8 \pi^2 R^6}{3Ic^3}
\label{eq:k_pulsar}
\end{equation}
in cgs units, where we assume the parameters of the canonical pulsar of radius $R$ 
and moment of inertia $I$, assumed to be $10^6$~cm and 
$10^{45}\mathrm{~g~cm}^2$ respectively. Equation~\ref{eq:p_t_ridley} is used to calculate
a value for $P$ at time $t$, and then the period derivative is calculated as 

\begin{equation}
    P^{n-2} \dot{P} = kB^2 \sin^2 \chi,
\label{eq:p_dot}
\end{equation}
where, for models with time-varying $\chi$, we use $\chi(t)$ instead.
The so-called ``death line'' which demarcates the area of the $P$-$\dot{P}$ diagram
in which few radio pulsars are observed. \citet{bhattacharya1992} described this 
line as

\begin{equation}
    P_\mathrm{death} = \sqrt {\frac{B}{1.7 \times 10^{11}\mathrm{~G}} },
\label{eq:fk06_deathline}
\end{equation}
and that pulsars with period $P>P_\mathrm{death}$ will be radio-quiet. 

\subsubsection*{CS06 Model}
For the pulsar spindown model of \citet{cs06}, the pulsar death line is represented by
the equation

\begin{equation}
    P_\mathrm{death} = \left[ 0.81 \times \left(\frac{B}{10^{12}\mathrm{~G}} \right) 
        \left( \frac{1\mathrm{~s}}{P_0}\right) 
                        \right]^\frac{2}{n+1}
    \label{eq:cs06_deathline}
\end{equation}
and, as with Equation~\ref{eq:fk06_deathline}, radio emission ceases when 
$P>P_\mathrm{death}$. However, $P$ and $\dot{P}$ are now calculated by integrating

\begin{eqnarray}
    \dot{P} & = & 3.3 \times 10^{-16} \left( \frac{P}{P_0}\right)^{2-n}
    \left( \frac{B}{10^{12}\mathrm{~G}} \right)^2
    \left( \frac{P_0}{1\mathrm{~s}} \right)^{-1} \nonumber \\
    & & \left( 1 - \frac{P}{P_\mathrm{death}} \cos^2 \chi \right)
    \label{eq:cs06_ppdot}
\end{eqnarray}
with respect to time, and then solving for the period.

\subsection{Evolving the pulsar through the Galactic potential}\label{sec:galpot}
Once a pulsar's time-evolved period and period derivative have been calculated, it
must be assigned a position in the Galaxy. Again, following FK06, 
initial positions are chosen in the following way:
\begin{enumerate}
    \item A radial position is chosen using the radial model of \citet{yk2004};
    \item The pulsar is positioned along one of the Galactic spiral arms, at the radius
        given in the previous step. The position is represented by $x$-$y$ coordinates
        in the plane of the Galaxy;
    \item The pulsar is assigned a third coordinate, $z$, perpendicular from the Galactic
        plane, using the same method as in \S~\ref{sec:scaleheight}, however this time
        using a scale height of only 50~pc.
\end{enumerate}

Pulsar birth velocities for each of the $x$, $y$ and $z$ directions are typically assigned
from a Gaussian distribution centered on $0\mathrm{~km~s}^{-1}$ with a width of 
$180\mathrm{~km~s}^{-1}$. The mean and standard deviation of this distribution may be
varied, and additional distributions are simple to implement. The pulsar is then evolved
from its initial position and velocity for a time equal to the age of the pulsar, using
the model by \citet{ci1987} as modified by \citet{kg1989}, giving a final position of 
the pulsar in the Galaxy. Models of the Galactic electron distribution may then be applied
as discussed in \S~\ref{sec:electrondist}.

\section{Generating a synthetic pulsar population}\label{sec:generatingpopulation}
All of the algorithms discussed in \S~\ref{sec:snapshot} are provided as 
stand-alone functions which can be imported to user-defined scripts. For `standard 
usage' of the software, however, command-line scripts are provided, which mimic the
behaviour of the \textsc{Psrpop} executables. These scripts form a pipeline for
creating a model population and applying different pulsar survey parameters to it.

The processes for generating synthetic populations using both the `snapshot' and 
evolutionary methods are discussed in \S~\ref{sec:populate} and \S~\ref{sec:evolve}. 
However, 
typical operation of both methods relies on our simulating pulsar survey sensitivity
thresholds. The method used to do this is discussed first, in \S~\ref{sec:dosurvey}.

\subsection{Simulating a pulsar survey}\label{sec:dosurvey}
Pulsars in the model population can be run through a series of survey parameters to
see if they would, in theory, be detected in such a survey. As we saw in the previous
section, this can be used to constrain the population based upon known detections, but
equally this could be used to make projections about future surveys with hypothetical
survey parameters.

\begin{figure}
	\begin{center}	
		\includegraphics[width=8.5cm]{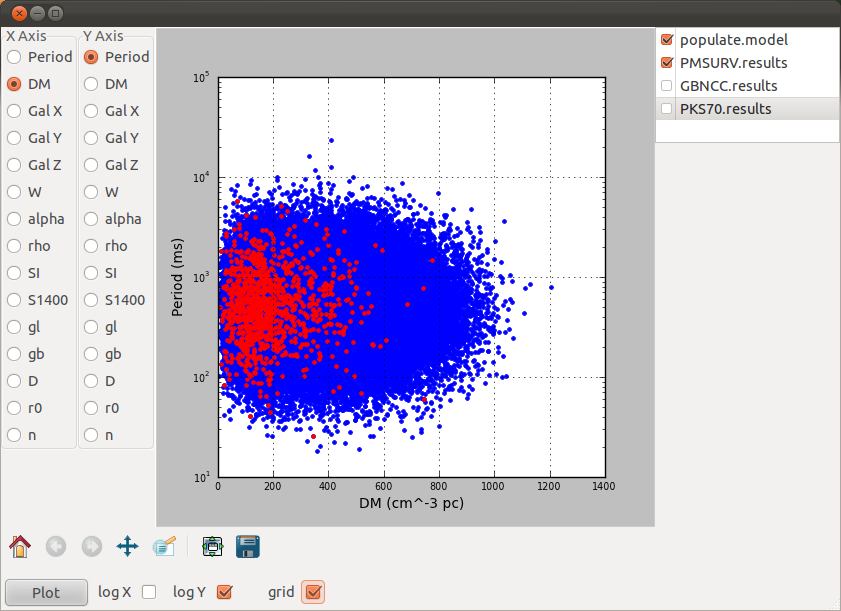}
	\end{center}
	\caption{Screenshot of the GUI for inspecting population models. Shown are
	    the whole model population generated by \textsc{PsrPopPy}, and 
		the subset detected in a simulated Parkes Multibeam survey.}
	\label{fig:wxView}
\end{figure}

\begin{table}
	\begin{center}
	\caption{Observational parameters required for simulating a pulsar survey
			using \textsc{PsrPopPy}. The examples are the Parkes southern pulsar survey 
			\citep[PKS70, ][]{mld+95}, the Parkes multi-beam pulsar survey \citep[PMPS, 
			][]{mlc+01} and the Parkes 6.5~GHz multi-beam pulsar survey 
			\citep[MMB,][]{bates2010}.}
		\begin{tabular}{lccc}
		\toprule
		& PKS70 & PMPS & MMB \\
		\midrule
		Degradation factor, $\beta$ & 1.2 & 1.2 & 1.2 \\
		Gain, $G$ ($\mathrm{K~Jy}^{-1}$) & 0.64 & $\sim 0.7$ & 0.6 \\
		Integration time, $t_\mathrm{obs}$ (s) & 157.3 & 2100 & 1055 \\
		Sampling interval, $t_\mathrm{samp}$ ($\mu$s) & 300 & 250 & 125\\
		System Temperature, $T_\mathrm{sys}$ (K) & 35 & 25 & 40\\
		Centre frequency, $f$ (MHz) & 436 & 1352 & 6591\\
		Bandwidth, BW (MHz) & 32 & 288 & 576 \\
		Channel width, $\Delta f$ (MHz) & 0.125 & 3 & 3\\
		Number of polarizations, $n_\mathrm{p}$  & 2 & 2 & 2 \\
		\\
		Beam FWHM (arcmin) & 45 & 14 & 3.2 \\
		\\
		Min. RA ($\degree$) & 0 & 0 & 0 \\
		Max. RA ($\degree$) & 360 & 360 & 360 \\
		Min. Dec ($\degree$) & $-90$ & $-90$ & $-90$ \\
		Max. Dec ($\degree$) & 0 & +90 & $+90$\\
		\\
		Min. Galactic longitude ($\degree$)& $-180$ & $-150$ & $-60$ \\
		Max. Galactic longitude ($\degree$)& $+180$ & $+50$ & $+30$ \\
		Min.\@Galactic latitude ($\degree$) & $-90$ & $-6$ & $-0.25$\\
		Max. Galactic latitude ($\degree$)& $+90$ & $+6$ & $+0.25$ \\
		\\
		Completed fraction & 1.0 & 1.0 & 1.0 \\
		Detection S/N & 9.0 & 9.0 & 9.0 \\
		\bottomrule
		\end{tabular}
		\label{table:surveyfile}
	\end{center}
\end{table}

\begin{table}
\begin{center}
\caption{Parameters used to simulate the pulsar population in \S~\ref{sec:lumsims}.}
    \begin{tabular}{lr}
    \toprule
     Radial distribution model & \citet{lfl+06} \\
     Initial Galactic $z$-scale height & 50 pc \\
     \\
     Luminosity distribution & Log-normal \\
     $\langle{\log_{10}\,\mathrm{L~(mJy~kpc^2)}}\rangle$ & $-1.1$\\
     $\mathrm{std}(\log_{10}\,\mathrm{L~(mJy~kpc^2)}))$ & $0.9$\\
     \\
     Spectral index destribution & Gaussian \\
     $\langle{\alpha}\rangle$ & $-1.4$ \\
     $\mathrm{std}(\alpha)$ & 0.96 \\
     \\
     Initial period distribution & Gaussian\\
     $\langle{\mathrm{P~(ms)}}\rangle$ & $300$\\
     $\mathrm{std}(\mathrm{P~(ms)})$ & $150$\\
     \\
     Pulsar spin-down model & FK06\\
     Beam alignment model & Orthogonal \\
     Braking index & 3.0\\
     Max pulsar age & 1~Gyr\\
     Initial $B$ field ditribution & Log-normal\\
     $\langle{\log_{10}B\,\mathrm{~(G)}}\rangle$ & $12.65$\\
     $\mathrm{std}(\log_{10}B\,\mathrm{~(G))}))$ & $0.55$\\
     \\
     Scattering model & \citet{bcc+04}\\
     \\
     Number of detectable pulsars & \multirow{2}{*}{1206}\\
     in the PMPS \& SWIN surveys \\
    \bottomrule
    \end{tabular}
    \label{table:sims}
\end{center}
\end{table}

\subsubsection*{Describing survey parameters}
Surveys parameters are defined in plain text files, making it easy for users to add
or edit their own surveys. Examples of the parameters used in these `survey files' 
are shown in Table~\ref{table:surveyfile}, and can then be used to predict whether 
the survey would be able to detect each of the pulsars.

For added precision, instead of describing the survey region by the bounds in either 
Galactic or equatorial coordinates, it is possible to provide a list of pointing 
coordinates (in either coordinate system), with associated gain and observation length
values, for the survey. This may be useful, for example,
for drift-scan surveys which are not easily described by a bounding region or for 
on-going surveys which could also be described, though less precisely, using a very low 
completed fraction, or for multibeam surveys which have highly variable gain values
away from the central beam.

\begin{figure*}
\begin{center}	
    \includegraphics[width=16cm]{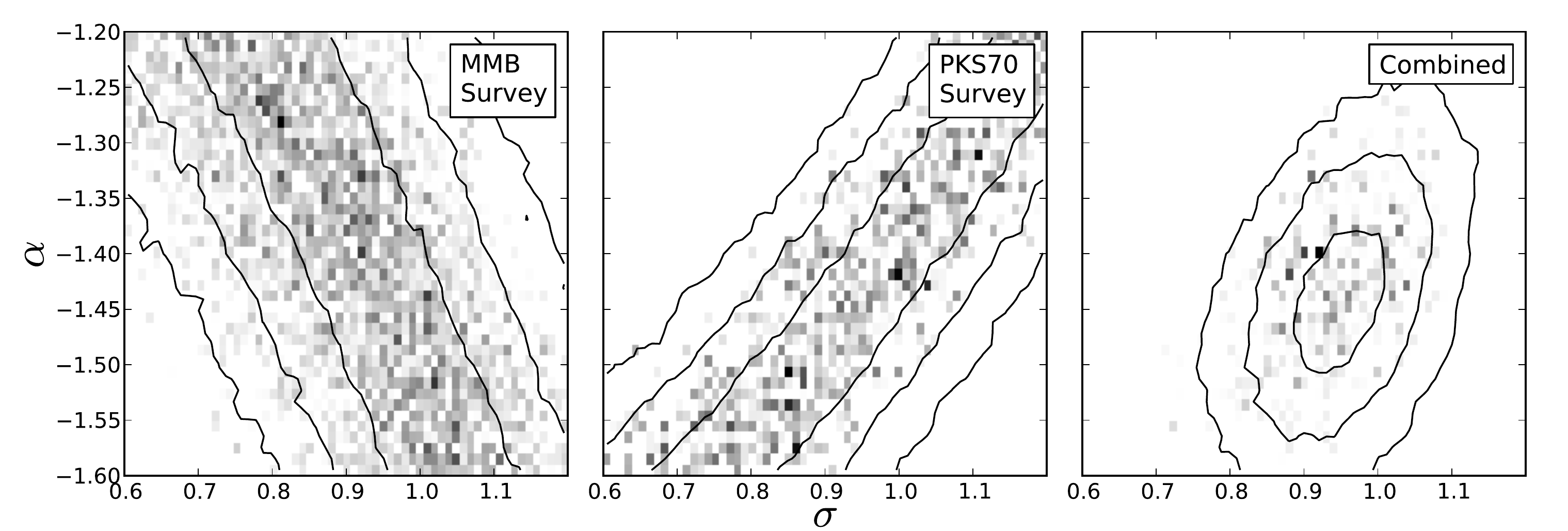}
\end{center}
\caption{ 
    Grey-scale plots of likelihoods overlaid with confidence-level contour lines
    \citep[as defined in][]{blv13} for population simulations with varying spectral
index parameters $\alpha$ and $\sigma$ (see Section~\ref{sec:specindexintro}).}
\label{fig:spectralindexsims}
\end{figure*}

\subsubsection*{Detection thresholds}\label{sec:detection}
The first test to be done is whether the pulsar is inside the region described by the 
RA/Dec or $l$/$b$ ranges in the text file. Any pulsars which are inside
this region then have their theoretical signal-to-noise ratio 
\begin{equation}
\mathrm{S/N} = \frac{S_\nu G \sqrt{n_\mathrm{p} t_\mathrm{obs} \Delta f }}{\beta T_\mathrm{tot}} \times \sqrt{\frac{1-\delta}{\delta}}
\label{eq:snr}
\end{equation}
where total temperature is the sum of the system, sky and cosmic microwave
background (CMB) temperatures,
\begin{equation}
    T_\mathrm{tot} = T_\mathrm{sys} + T_\mathrm{sky} + T_\mathrm{CMB},
\end{equation}
and many of the other terms are defined in Table~\ref{table:surveyfile}. The pulse
duty cycle, 
\begin{equation}
\delta = W_\mathrm{eff}/P,
\end{equation}
where $W_\mathrm{eff}$ is the effective pulse width and $P$ the pulse period.

The effective pulse width is not the intrinsic width of pulses from the pulsar. It is 
given by
\begin{equation}
W_\mathrm{eff} = \sqrt{W_\mathrm{int}^2 + t_\mathrm{samp}^2 + \Delta t^2 + \tau_\mathrm{s}^2}
\label{eq:weff}
\end{equation}
where $W_\mathrm{int}$ is the intrinsic pulse width, $t_\mathrm{samp}$ is the sampling
time of the hardware used to record survey data, and
\begin{equation}
\Delta t = 8.3 \times 10^6~\mathrm{ms} \times \mathrm{DM} \times \frac
            {\Delta f_\mathrm{MHz}}{f_\mathrm{MHz}^3}.
\end{equation}
is the dispersive smearing across a single frequency channel of bandwidth
$\Delta f_\mathrm{MHz}$ at frequency $f_\mathrm{MHz}$, both in units of MHz.

The final term in Equation~\ref{eq:weff} is the pulse smearing due to scattering by
free electrons in the interstellar medium, $\tau_\mathrm{s}$. For this, we use the
empirical scattering fit of \citet{bcc+04},
\begin{eqnarray}
    \log_{10}\tau_\mathrm{s} &=& -6.46+0.154\log_{10}\mathrm{DM}
   +1.07(\log_{10}\mathrm{DM})^2 \nonumber \\
   & & -3.86\log_{10}{f_\mathrm{GHz}},
\label{bhatscatter}
\end{eqnarray}
for a pulsar with dispersion measure DM, and where frequency $f_\mathrm{GHz}$ 
is now given in GHz. Since there is a large 
scatter about this relationship, we pick the final value of $\tau_\mathrm{s}$ from a 
Gaussian distribution centred on the value computed from Equation~\ref{bhatscatter}. To 
enable investigations of this scattering relationship, users are also able to use
custom values for the frequency coefficient in this equation.

During a pulsar survey where the sky is `tiled' with observations, pulsars are not 
commonly found at the centre of the survey beam, rather, they are offset by some angle
which causes them to be detected with a lower S/N than if they were positioned at the
beam centre. There are two methods incorporated in \textsc{PsrPopPy} to reproduce this
behaviour. In the most simple model, following \citet{lorimer1993}, a Gaussian telescope
beam is assumed, where the gain, $G$ at any offset, $r$, from the beam centre is given
by
\begin{equation}
    G = G_0 \exp \left( \frac{-2.77r^2}{w^2} \right)
\label{eq:gainsimple}
\end{equation}
for a telescope with FWHM $w$ and where the gain at the centre of the beam is $G_0$.
The square of the offset is chosen from a pseudorandom uniform distribution, 
$0 \leq r^2 \leq w^2/4$, so as to give uniform coverage over the area of the survey beam.

An Airy disk model, providing improved precision, is also included,
\begin{equation}
G=G_0 \bigg(\frac{2{\rm J_1}(k a \sin r)}{k a \sin r}\bigg)^2
\label{eq:gainairy}
\end{equation}
where $J_1$ is a Bessel function of the first kind with index 1, 
$k = 2\pi/\lambda$ is the wavenumber
for an observing wavelength $\lambda$, and $a$ is the effective aperture radius. This
is slightly slower, but may be useful in certain situations --- for example the PALFA
survey at Arecibo, where the first sidelobe is roughly as sensitive at the main beam 
of the Parkes radio telescope (see Swiggum et al.\,in press).

More accurate values of the modified gain can be obtained by using a list of survey
beam positions and, in the case of multibeam surveys, the value of $G_0$ for each 
observation. From this list, the offset from each pulsar to the nearest survey beam
is calculated and then used in one of Equations~\ref{eq:gainsimple} or 
\ref{eq:gainairy}.

Using these relationships, the value of S/N can then be computed using 
Equation~\ref{eq:snr}, and if it is greater
than the detection threshold (see Table~\ref{table:surveyfile}), then the pulsar is
counted as detected by the survey. \textsc{PsrPopPy} also keeps track of how many pulsars
are outside the survey region, are smeared out completely (that is, $W_\mathrm{eff} > P$) 
or are simply too faint to be detected. Results are then reported and can be stored in 
a text file. A new population model (in the same format as that generated in 
\S~\ref{sec:populate}) is also written to disk, for each survey. This allows the results 
to be directly compared to one another.

If multiple pulsar surveys are used, these same numbers are computed for each survey,
and reported individually. \textsc{PsrPopPy} also records the number of discoveries
(not only detections) in each survey. This allows the potential of future surveys to be
more carefully calculated.

\subsection{Populating the model galaxy with POPULATE}\label{sec:populate}
The most basic method for creating a model population is as follows
\begin{enumerate}
  \item the user selects a number of pulsars to be generated;
  \item for each simulated pulsar in turn, values for each of the pulsar parameters
  	are drawn from the user-specified (or default) distributions.
\end{enumerate}
This method might be suitable for situations where a Galactic population of $X$ pulsars
is hypothesised, which could then be tested from simulating survey results,
or other means.

Commonly, users wish to generate a population based upon constraints provided by the
large-scale pulsar surveys which have been performed to date. In this case, the user 
can provide a list of surveys they wish to use to constrain the model, and the total 
number of pulsars, $n$, that should be detected in these surveys. Then, the method 
differs slightly.

\begin{enumerate}
  \item the code continually generates new synthetic pulsars;
  \item for each simulated pulsar in turn, values for each of the pulsar parameters
  	are drawn from the user-specified (or default) distributions;
  \item each pulsar is run through each of the simulated surveys in turn (see 
    \S~\ref{sec:dosurvey});
  \item if the pulsar is detected in any of the surveys, a counter is incremented and a
      new pulsar generated;
  \item if the pulsar is not detected, it remains in the model, but the counter is not
    incremented;
  \item when the counter reaches $n$, the loop terminates.
\end{enumerate}

\begin{figure}
\begin{center}	
    \includegraphics[width=8.5cm]{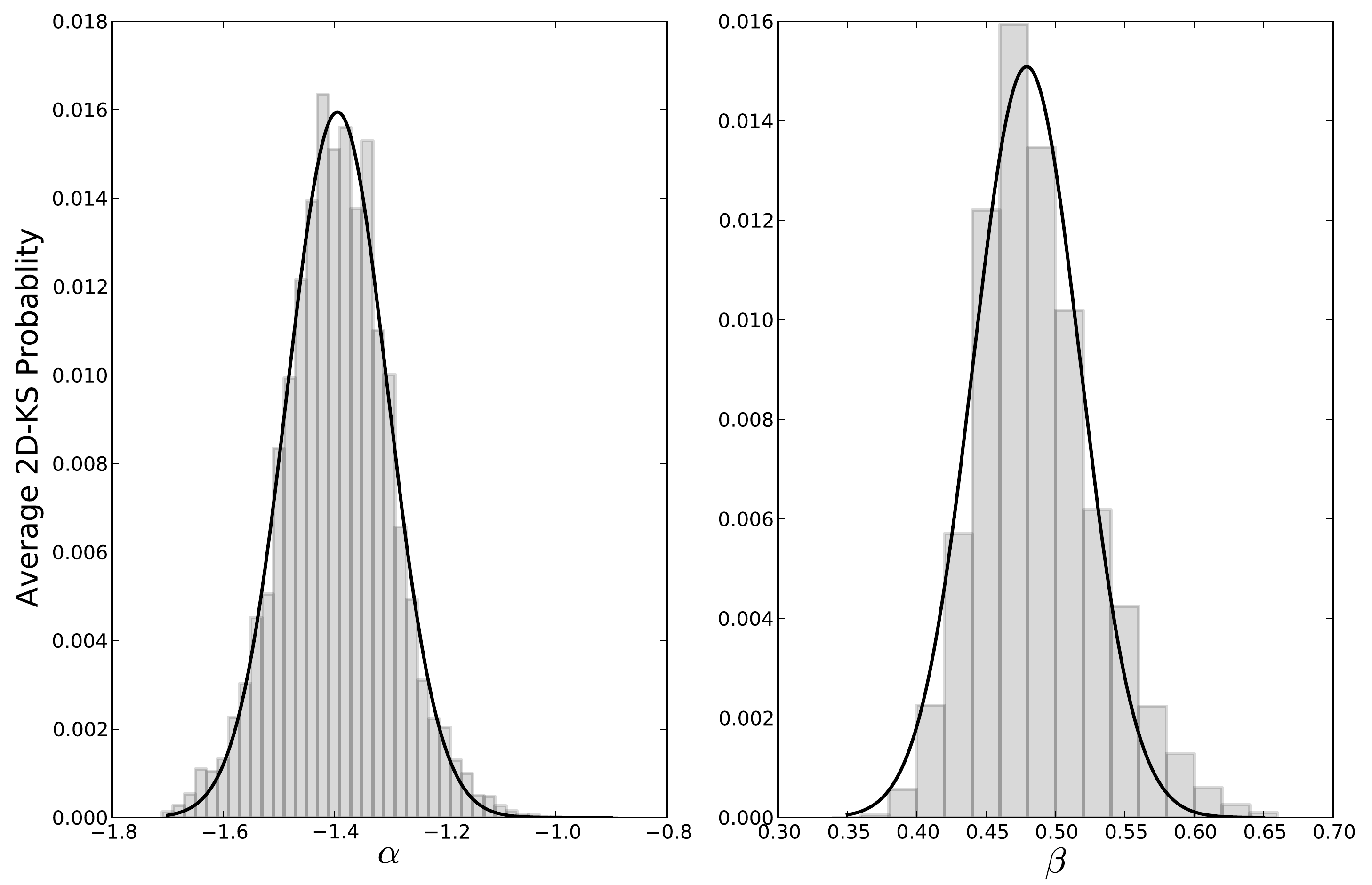}
\end{center}
\caption{Marginalized probability density functions obtained for $\alpha$ (left) and $
        \beta$ (right). Best fits are shown with a solid line, giving the results 
        $\bar{\alpha} = -1.4$, $\bar{\beta} = 0.48$.
}
\label{fig:alphabetahist}
\end{figure}

\begin{figure*}
\begin{center}	
    \includegraphics[width=17cm]{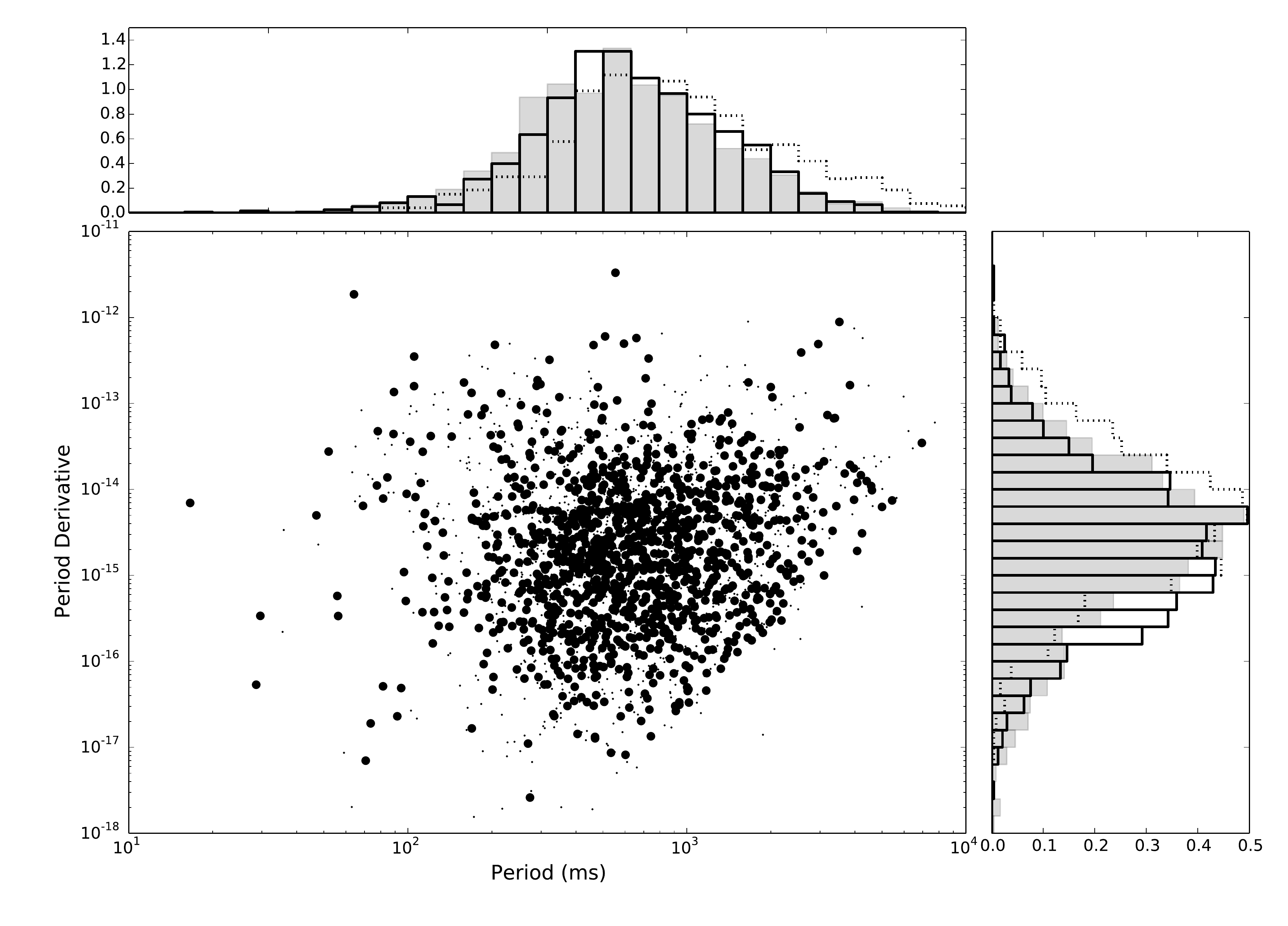}
\end{center}
\caption{$P$-$\dot{P}$ diagram of pulsars detected in model PMSURV and SWIN surveys
    produced using \textsc{evolve} and
    the luminosity distribution 
    parameters $\alpha = -1.4$ and $\beta = 0.48$ (bold points).
    $P$-$\dot{P}$ values that were taken from the pulsar catalogue are shown 
    for comparison (smaller points). 
    Histograms of the $P$ and $\dot{P}$ distributions are also shown, with grey bars 
    for catalogue sources, and with solid lines for the simulated pulsars. For 
    comparison, step histograms (dotted) are shown for a simulation
    using the values $\alpha=-1.0, \beta=0.7$.
}
\label{fig:ppdot}
\end{figure*}

\subsection{Simulating pulsar period derivatives with EVOLVE}\label{sec:evolve}
For evolutionary simulations of the pulsar population, \textsc{PsrPopPy} includes 
models for generating period derivatives for each pulsar, based upon work by 
FK06 and \citet{cs06}. The method follows that discussed by \citet{ridley2010}
and includes a variety of pulsar beaming models and the option of including a decay in 
the angle between the spin and magnetic axes of each pulsar. 

Although the evolution code is contained in a separate executable, the output file 
format is identical to that used in the rest of \textsc{PsrPopPy}, and so is completely
compatible with the output from the snapshot simulations. The \textsc{Evolve} code 
is outlined below;

\begin{enumerate}
  \item the code continually generates new synthetic pulsars as in
        \S~\ref{sec:populate};
  \item for each simulated pulsar in turn, the rotation period is drawn from the 
      chosen distribution, the age, $\tau_\mathrm{init}$, of the pulsar is chosen
      from a flat ditribution between $0$ and a maximum age, $t_\mathrm{max}$
      (corresponding to the age of the Galaxy), and a 
      magnetic field is chosen from a log-normal distribution;
  \item an alignment angle between the spin and magnetic axes is chosen, and a decay
      timescale for this alignment can be chosen, if required;
  \item the pulsar is assigned a braking index, and then one of the two spindown models
      is applied. If the user wants to apply a `deathline', the pulsar is evaluated
      to see if it has crossed into the `dead' region;
  \item each `live' pulsar is randomly assigned a birth position in Galactic plane 
      and a birth velocity;
  \item the pulsar is then evolved through the Galactic potential for a time equal to 
      the age of the pulsar;
  \item the pulsar is assigned a radio luminosity and spectral index, and run through 
      any selected model surveys, as discussed in \S~\ref{sec:dosurvey};
  \item if the pulsar is detected in any of the surveys, a counter is incremented and a
    new pulsar generated;
  \item if the pulsar is not detected, it remains in the model, but the counter is not
    incremented;
  \item when the counter reaches the desired number of detections, the loop terminates.
\end{enumerate}

\subsection{Running model populations through further surveys}\label{sec:dosurvey}
Once a population model has been created, with the evolutionary or snapshot method,
a common use of the model is to run alternate pulsar surveys over the model --- either
in order to test the input distributions in some way (for example, using surveys
at multiple frequencies to constrain spectral evolution), or to make predictions of
how many pulsars you might expect future surveys to detect or discover. The script 
\textsc{dosurvey} is provided for exactly this purpose, which performs the following
steps:

\begin{enumerate}
    \item a specified population model is read in;
    \item survey models are created from text files outlining the survey parameters. 
        Multiple surveys may be selected;
    \item for each survey, in turn, the pulsars in the population model are run 
        through the equations listed in \S~\ref{sec:detection} to calculate 
        the pulsar's S/N;
    \item the code totals not only the number of pulsars detected in each survey, but
        also how many pulsars were first detected by that survey --- i.e.\,giving an 
        estimate of the number of discoveries;
    \item a smaller population model, containing only the detected pulsars, is then
        written to disk for each survey.
\end{enumerate}

\subsection{Analysing population models}
Two scripts have been developed as part of \textsc{PsrPopPy} to allow users to visually
inspect the population models generated by the processes discussed in 
\S~\ref{sec:populate}, \S~\ref{sec:evolve}, and \S~\ref{sec:dosurvey}.

The first tool will plot simple histograms of all the pulsars in the model population, 
using any given parameter (for example, Galactic longitude/latitude, period, DM, etc.\@)

Secondly, \textsc{PsrPopPy} provides an interactive graphical user interface for 
plotting pulsar parameters against one another (for a screenshot see 
Figure~\ref{fig:wxView}). Users can select models in the current working directory
to be plotted, select which parameters to plot, and choose to plot using linear or
logarithmic $x$ and $y$ axes. Plots can also be saved in various file formats, 
including vector formats such as postscript.

\subsection{Combining PsrPopPy functions}
\textsc{PsrPopPy} has been designed to allow user-generated scripts to be simple to
implement. All population models are written in a uniform format and, in fact, are 
simply serialised objects from the code. This means that no code is required to 
parse models back into memory, and that high-level scripts can be used to pass a model
directly between codes, without having to write to disk and then read it out again. 
This also removes the need to modify the codebase whenever additional parameters are 
added to the models, making the code more robust. 

\section{Applications of \textsc{PsrPopPy}}\label{sec:applications}

\subsection{Spectral Index Distribution}\label{sec:specindexdist}
In order to test the ability of \textsc{PsrPopPy} to reproduce results previously 
obtained with \textsc{PsrPop}, we first ran simulations following 
\citet[][see \S~4 of that paper for details of the method]{blv13},
to determine the spectral index
distribution of normal pulsars. The only difference allowed was to perform only 50
realisations per bin instead of 500. We are able to reproduce the Figure~1 of 
\citeauthor{blv13}, shown here in Figure~\ref{fig:spectralindexsims}, which in turn would
lead to the same conclusions reached in that work --- the best-fitting spectral index
distribution from these simulations is $\alpha=-1.4$, $\beta = 1.0$.

\subsection{Modelling the pulsar luminosity distribution}\label{sec:lumsims}
As discussed in \S~\ref{sec:lumlaw}, the pulsar luminosity distribution is sometimes 
parameterised in terms of the period, $P$, and period derivative, $\dot{P}$, using the
formulation show in Equation~\ref{eq:ppdotlum}. In this section, we use the 
evolutionary simulation code \textsc{evolve} to estimate the value of the parameters
$\alpha$ and $\beta$ in this equation. We kept value of $\gamma$ in this equation fixed
at $0.18 \mathrm{~mJy~kpc}^2$, the optimal value according to FK06.

For values of $\alpha$ and $\beta$ covering a grid covering the range
$-1.6 < \alpha <0.8$ and $0.3 < \beta < 0.7$, model pulsar populations were generated
using \textsc{evolve} as per the specifications outlined in Table~\ref{table:sims}. The 
populations were grown until 1206 pulsars per population were detected in simulations of the
Parkes multibeam pulsar survey \citep[PMSURV, ][]{mld+95} and the two Swinburne pulsar 
surveys at
higher Galactic latitudes \citep{ebsb01,jbo+09}. This is the number of normal pulsars 
(defined as $P>30\mathrm{~ms}$, $\dot{P}<1\times10^{-12}$) listed in the pulsar
catalogue \citep{mhth05} as detected by any one or more of these three surveys.

For each grid point $(\alpha, \beta)$, 25 realisations of the simulation were performed.
For each, the $P$ and $\dot{P}$ values of the resulting population model were compared 
to those of the 1206 pulsars listed in the pulsar
catalogue using the 2-dimensional Kolmogorov-Smirnov test \citep{pftv86}.
While the probabilities
generated by this algorithm are not statistically 
rigorous, this test does provide a helpful way of quantifying how well two 2-d
distributions match one another. The average value of this probability over the 25 
realisations was stored, as were the models generated by \textsc{dosurvey}.

Marginalising over $\alpha$ and then $\beta$ in turn, by averaging the probabilities for 
each $\alpha$ or $\beta$ value, the distributions shown in Figure~\ref{fig:alphabetahist}
were obtained. Fitting Gaussian functions to these probability distributions, we obtained
best-fit values of $\bar{\alpha} = -1.39 \pm 0.09$ and $\bar{\beta} = 0.48 \pm 0.04$. An
example $P$-$\dot{P}$ diagram generated using these values for $\alpha$ and $\beta$ is
then shown in Figure~\ref{fig:ppdot}. Histograms are also shown for a simulation using 
the values $\alpha=-1.0, \beta=0.7$, and are clearly skewed to longer periods and 
higher values of the period derivative.

It is worth noting that while we have derived values of $\alpha$ and $\beta$ by 
comparing our simulations to the known pulsar population, we have obtained a very 
similar result to \citet{perera2013}. Their analysis considered the relationship for
gamma-ray luminosities, $P$ and $\dot{P}$, and derived values of 
$\alpha = -1.36 \pm 0.03$ and $\beta = 0.44 \pm 0.02$. This one-to-one correspondence
between empirical luminosities is remarkable given that the fraction of the 
spin-down energy loss going into gamma-ray emission is substantially greater than 
in the radio. The X-ray emission does not scale in a similar
manner for rotation-powered pulsars \citep[e.g.][]{possenti2002, kargaltsev2012}.
Further theoretical studies to explain these results are definitely warranted, and 
we plan to explore this further in future work.

\begin{figure}
\begin{center}	
    \includegraphics[width=8.5cm]{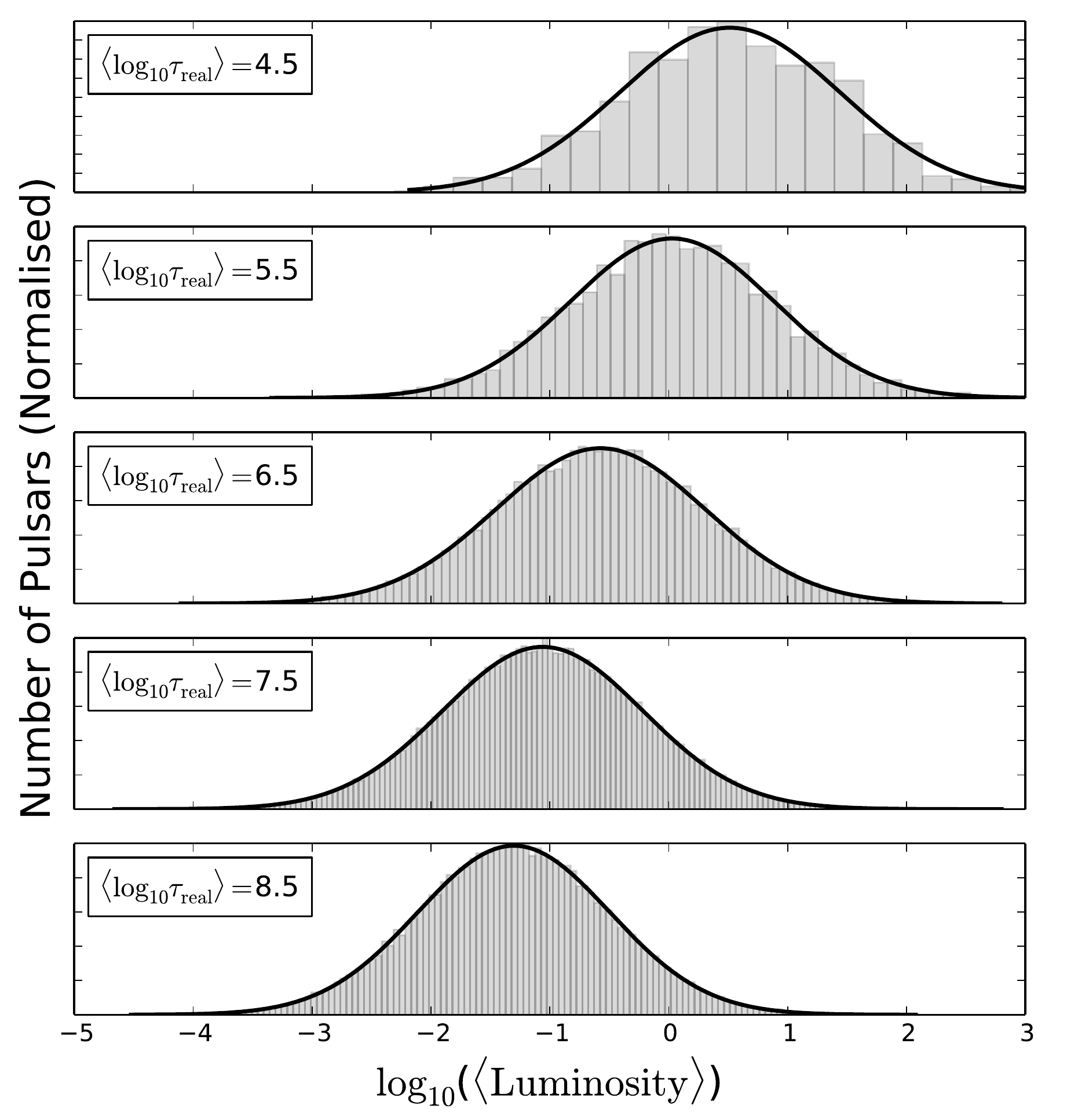}
\end{center}
\caption{Histograms of pulsar luminosities for each `real age' bin. Fitted Gaussians
        are shown with a solid black line.
        }
\label{fig:lumhistsreal}
\end{figure}

\begin{figure}
\begin{center}	
    \includegraphics[width=8.5cm]{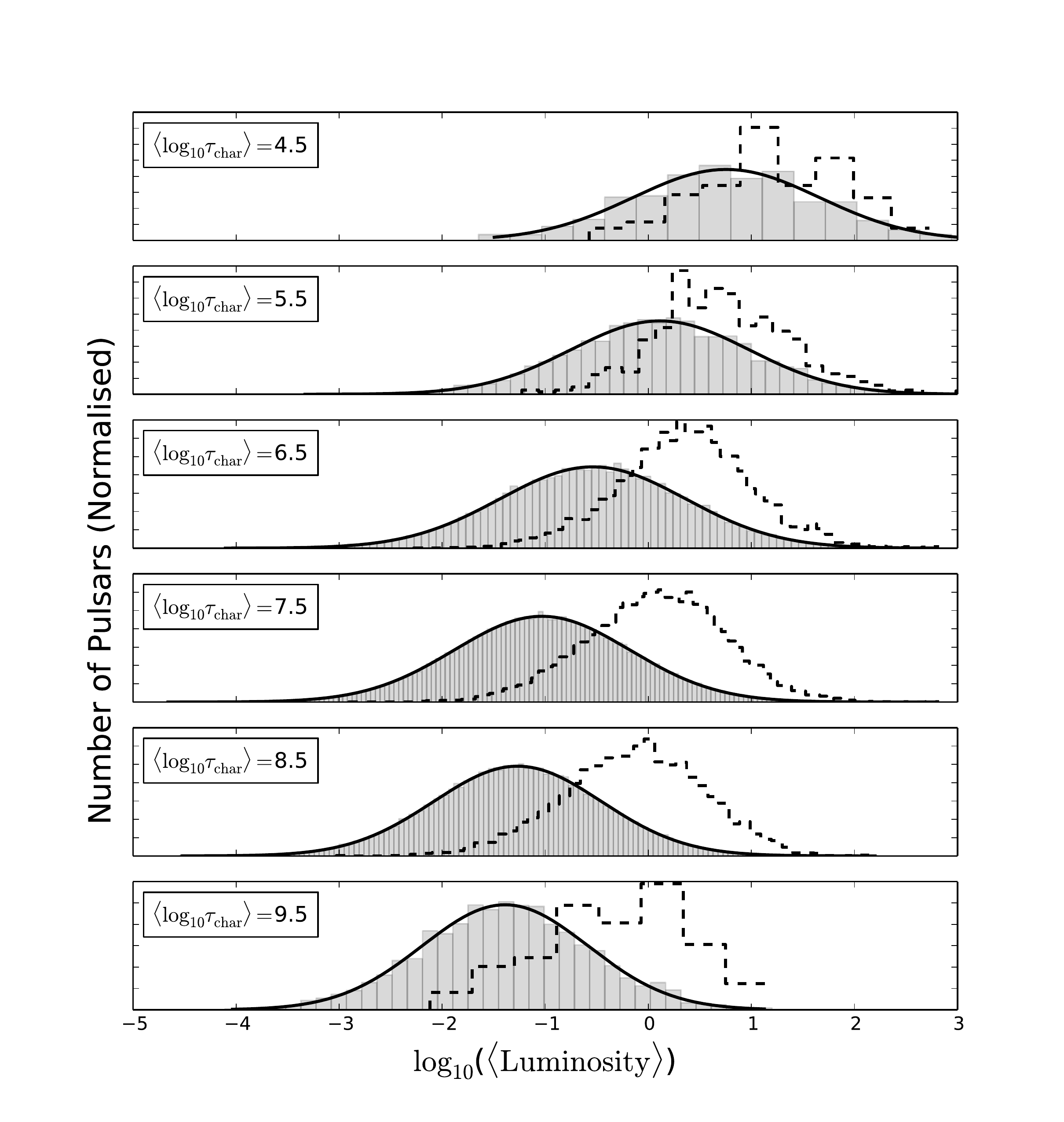}
\end{center}
\caption{Normalised histograms (grey bars) of pulsar luminosities for 
        each characteristic age bin. 
        Fitted Gaussians are shown with a solid black line. The dotted lines show
        normalised histograms of the pulsars detected in a model all-sky survey 
        with the SKA.
        }
\label{fig:lumhistschar}
\end{figure}

\subsection{Scaling laws for the underlying population}\label{sec:scaling}
Having, in the previous section, constrained the values of $\alpha$ and $\beta$
in Equation~\ref{eq:ppdotlum}, we then should be able to see how pulsar luminosities
vary with age (both the `real' age, $\tau_\mathrm{real}$ in our simulations, 
and the characteristic age, given by $\tau_\mathrm{char} = P/2\dot{P}$). 

\begin{figure}
\begin{center}	
    \includegraphics[width=8.5cm]{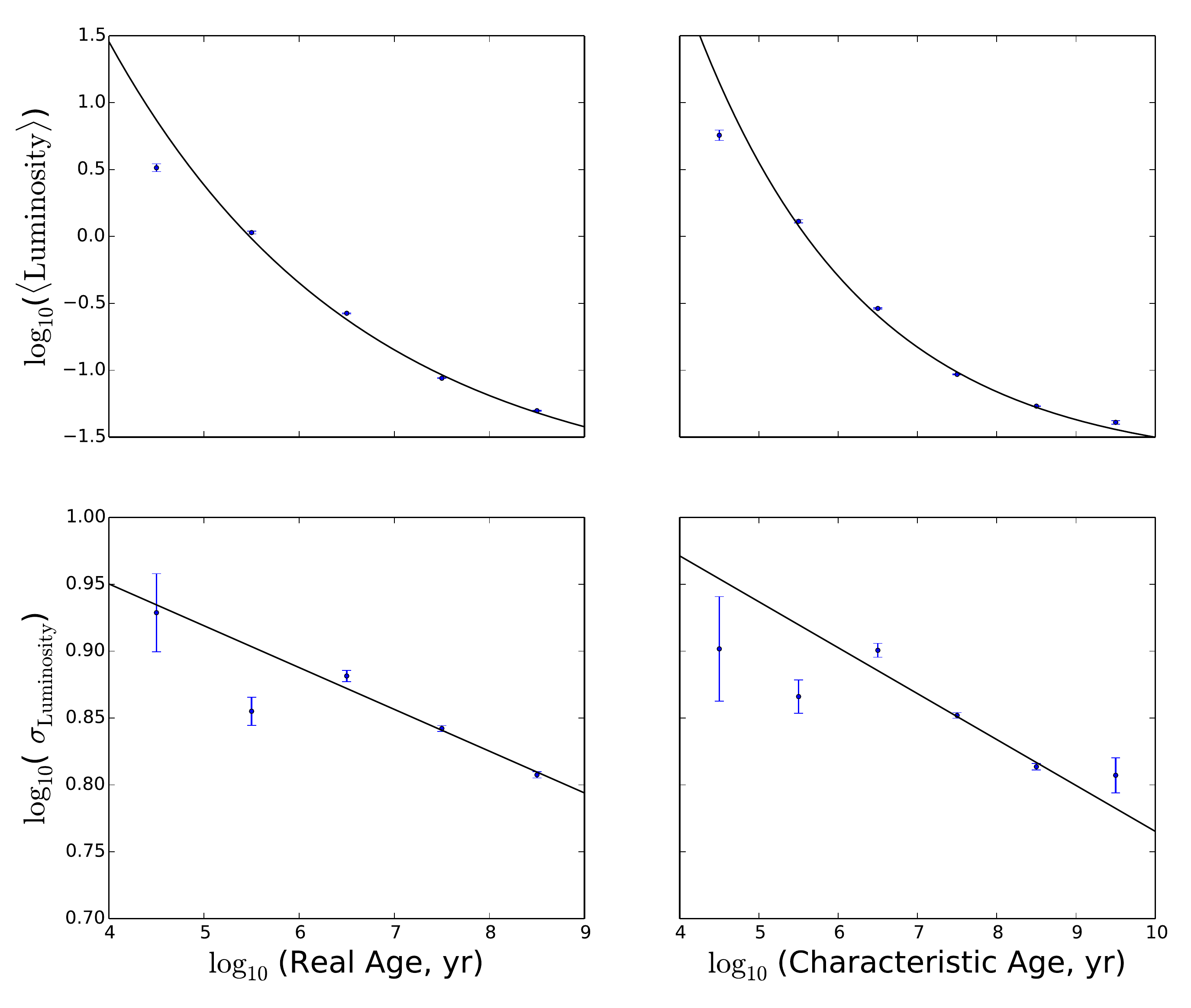}
\end{center}
\caption{Mean (upper) and standard deviation (lower) values, with associated
    errors, from the fits performed in
    Figures~\ref{fig:lumhistsreal} (left) and \ref{fig:lumhistschar} (right). 
    Upper panels are fit by an
    exponential decay, the lower panels with a linear decay.
}
\label{fig:lumhistfits}
\end{figure}

First, a model
was created using the best values $\alpha=-1.39$, $\beta = 0.48$. Using only pulsars in 
the population which lie above the deathline (and are, therefore, considered to be
emitting radiation), the population was then divided into logarithmic bins in both
real and characteristic ages (with centre values $\log_{10} \tau = 4.5, 5.5 ..., 9.5$ 
for $\tau$ in years). For each of the age bins, histograms were made of pulsar 
luminosity (shown in Figures~\ref{fig:lumhistsreal} and \ref{fig:lumhistschar}). 
The histograms were fitted
with Gaussians of mean, $\mu$ and standard deviation $\sigma$, and the fitted values
of $\mu$ and $\sigma$, with the errors on these values taken from the resultant 
covariance matrix, are then plotted in Figure~\ref{fig:lumhistfits}.

The results from Figure~\ref{fig:lumhistfits} are that we can characterise the change
in pulsar luminosity, $L$, with the logarithm of age (both real age and
characteristic age) as an exponential decay, 
\begin{equation}
    \log_{10}L \propto \exp\left(- \frac{\log_{10}\tau}{\log_{10}\tau_\mathrm{decay}}\right) + a.
    \label{eq:ageexp}
\end{equation}
We obtained best fits of $\log\tau_\mathrm{decay} = 2.6 \pm 0.8$, $a = -1.9 \pm 0.3$ when
considering real ages, and $\log\tau_\mathrm{decay} = 2.1 \pm 0.4$, $a = -1.7 \pm 0.2$
for the characteristic ages.

The width of the luminosity distributions is also seen to change, shown in the lower 
panels of Figure~\ref{fig:lumhistfits}. We model this change as 
\begin{equation}
    \log_{10}\sigma_L = b \log_{10}\tau + c.
    \label{eq:ageexp}
\end{equation}
We obtained best fitting values of $b = -0.031 \pm 0.006, c = 1.08 \pm 0.05$ 
for pulsar real ages, and $b = -0.034 \pm 0.006, c = 1.11 \pm 0.05$ in the case
of characteristic ages.

Our simulations show, however, that the currently-known pulsar population
(simulated using a sub-sample detected by PMSURV and the Swinburne pulsar surveys)
does not contain enough low-luminosity pulsars
 to be able to measure this age-luminosity relationship. However, 
as shown in Figure~\ref{fig:lumhistschar}, the large number of pulsars expected to be
discovered in all-sky surveys using the SKA \citep[see, e.g.,][]{smits2008}, may 
begin to probe this relationship.

\section{Conclusions}

We have developed a software packge for the simulation of the Galactic pulsar
population. The package has been written mainly in Python, with reliance on some
older Fortran code for complex calculations. Using this software, 
we have demonstrated possible applications of the code:

\begin{itemize}
    \item in \S~\ref{sec:specindexdist} we reproduced the results of \citet{blv13}, who 
        characterised the spectral index distribution of normal pulsars as a Gaussian
        with mean $-1.4$ and standard deviation $1.0$;
    \item in \S~\ref{sec:lumsims}, the pulsar luminosity distribution was characterised
        in terms of the pulsar period, $P$, and period derivative, $\dot{P}$. We 
        obtained best-fit values for the power-law indices $\alpha$ and $\beta$ of
        $\alpha = -1.39 \pm 0.09$ and $\beta = 0.48 \pm 0.04$;
    \item in \S~\ref{sec:scaling} we looked at how the pulsar luminosity distribution
        varies with pulsar ages (both real and characteristic). It was seen that, 
        regardless of whether real or characteristic age is used, $\log_{10}L$ falls
        exponentially with $\log_{10}\tau$ with a decay constant of 2.1--2.6.
\end{itemize}

Further functionality will be added to the \textsc{PsrPopPy} code in the future, 
including: 
\begin{itemize}

    \item an improved treatment of millisecond pulsars, including models of binary
        parameters;
    \item using these binary parameters to model the effect of binary motion on 
        signal strength in pulsar surveys \citep[e.g.\,][]{blw13};
    \item the inclusion of pulse intensity distributions, and the extension of these
        to include rotating radio transients (RRATs);
    \item model the effects of scintillation on low-DM
        pulsars \citep[e.g.\,][]{cc97}.
\end{itemize}

Once some of these features have been implemented, it may become possible to repeat 
the simulations we have performed here, or others similar to them, instead focussing
on the millisecond pulsars, or RRATs. It may also be useful to include these pulsars
in predictions for future surveys. Of course, as the number of such known sources
increases, there will be better statistical models to be included in \textsc{PsrPopPy}, 
in turn increasing the accuracy of the software.

\section*{ACKNOWLEDGEMENTS}
The authors acknowledge support from WVEPSCoR in the form of a Research 
Challenge Grant. DRL is also supported by the Research Corporation for Scientific 
Advancement as a Cottrell Scholar.

We thank the referee for helpful comments on the manuscript.
\newpage

\bibliography{allrefs}

\begin{thebibliography}{}

\bibitem[\protect\citeauthoryear{{Bagchi}, {Lorimer}, \& {Wolfe}}{{Bagchi}
  et~al.}{2013}]{blw13}
{Bagchi} M., {Lorimer} D.~R.,  {Wolfe} S., 2013, MNRAS, 432, 1303

\bibitem[\protect\citeauthoryear{{Bates} et~al.}{{Bates}
  et~al.}{2011a}]{batesmsps}
{Bates} S.~D. et~al., 2011a, MNRAS, 416, 2455

\bibitem[\protect\citeauthoryear{{Bates} et~al.}{{Bates}
  et~al.}{2011b}]{bates2010}
{Bates} S.~D. et~al., 2011b, MNRAS, 411, 1575

\bibitem[\protect\citeauthoryear{{Bates}, {Lorimer}, \& {Verbiest}}{{Bates}
  et~al.}{2013}]{blv13}
{Bates} S.~D., {Lorimer} D.~R.,  {Verbiest} J.~P.~W., 2013, MNRAS, 431, 1352

\bibitem[\protect\citeauthoryear{{Bhat} et~al.}{{Bhat} et~al.}{2004}]{bcc+04}
{Bhat} N.~D.~R., {Cordes} J.~M., {Camilo} F., {Nice} D.~J.,  {Lorimer} D.~R.,
  2004, ApJ, 605, 759

\bibitem[\protect\citeauthoryear{{Bhattacharya} et~al.}{{Bhattacharya}
  et~al.}{1992}]{bhattacharya1992}
{Bhattacharya} D., {Wijers} R.~A.~M.~J., {Hartman} J.~W.,  {Verbunt} F., 1992,
  A\&A, 254, 198

\bibitem[\protect\citeauthoryear{{Boyles} et~al.}{{Boyles}
  et~al.}{2012}]{boyles2012}
{Boyles} J. et~al., 2012, ArXiv e-prints

\bibitem[\protect\citeauthoryear{{Burgay} et~al.}{{Burgay}
  et~al.}{2006}]{bjd+06}
{Burgay} M. et~al., 2006, MNRAS, 368, 283

\bibitem[\protect\citeauthoryear{{Carlberg} \& {Innanen}}{{Carlberg} \&
  {Innanen}}{1987}]{ci1987}
{Carlberg} R.~G.,  {Innanen} K.~A., 1987, AJ, 94, 666

\bibitem[\protect\citeauthoryear{{Contopoulos} \& {Spitkovsky}}{{Contopoulos}
  \& {Spitkovsky}}{2006}]{cs06}
{Contopoulos} I.,  {Spitkovsky} A., 2006, ApJ, 643, 1139

\bibitem[\protect\citeauthoryear{Cordes \& Chernoff}{Cordes \&
  Chernoff}{1997}]{cc97}
Cordes J.~M.,  Chernoff D.~F., 1997, ApJ, 482, 971

\bibitem[\protect\citeauthoryear{{Cordes} \& {Lazio}}{{Cordes} \&
  {Lazio}}{2002}]{ne2001}
{Cordes} J.~M.,  {Lazio} T.~J.~W., 2002, ArXiv Astrophysics e-prints
  (astro-ph/0207156)

\bibitem[\protect\citeauthoryear{{Edwards} et~al.}{{Edwards}
  et~al.}{2001}]{ebsb01}
{Edwards} R.~T., {Bailes} M., {van Straten} W.,  {Britton} M.~C., 2001, MNRAS,
  326, 358

\bibitem[\protect\citeauthoryear{Faucher-Gigu{\`e}re \&
  Kaspi}{Faucher-Gigu{\`e}re \& Kaspi}{2006}]{fk06}
Faucher-Gigu{\`e}re C.~A.,  Kaspi V.~M., 2006, ApJ, 643, 332, in press

\bibitem[\protect\citeauthoryear{{Gil}, {Gronkowski}, \& {Rudnicki}}{{Gil}
  et~al.}{1984}]{gil1984}
{Gil} J., {Gronkowski} P.,  {Rudnicki} W., 1984, A\&A, 132, 312

\bibitem[\protect\citeauthoryear{{Jacoby} et~al.}{{Jacoby}
  et~al.}{2009}]{jbo+09}
{Jacoby} B.~A., {Bailes} M., {Ord} S.~M., {Edwards} R.~T.,  {Kulkarni} S.~R.,
  2009, ApJ, 699, 2009

\bibitem[\protect\citeauthoryear{{Kargaltsev} et~al.}{{Kargaltsev}
  et~al.}{2012}]{kargaltsev2012}
{Kargaltsev} O., {Durant} M., {Pavlov} G.~G.,  {Garmire} G., 2012, ApJS, 201,
  37

\bibitem[\protect\citeauthoryear{{Keith} et~al.}{{Keith}
  et~al.}{2010}]{keith2010}
{Keith} M.~J. et~al., 2010, MNRAS, 409, 619

\bibitem[\protect\citeauthoryear{Kramer et~al.}{Kramer et~al.}{1998}]{kxc+98}
Kramer M., Xilouris K.~M., Lorimer D., Doroshenko O., Jessner A., Wielebinski
  R., Wolszczan A.,  Camilo F., 1998, ApJ, 501, 270

\bibitem[\protect\citeauthoryear{{Kuijken} \& {Gilmore}}{{Kuijken} \&
  {Gilmore}}{1989}]{kg1989}
{Kuijken} K.,  {Gilmore} G., 1989, MNRAS, 239, 651

\bibitem[\protect\citeauthoryear{{Levin} et~al.}{{Levin}
  et~al.}{2013}]{levin2013}
{Levin} L. et~al., 2013, MNRAS, 434, 1387

\bibitem[\protect\citeauthoryear{{Lorimer}}{{Lorimer}}{2012}]{lorimerIAU2012}
{Lorimer} D.~R., 2012, ArXiv e-prints

\bibitem[\protect\citeauthoryear{{Lorimer} et~al.}{{Lorimer}
  et~al.}{1993}]{lorimer1993}
{Lorimer} D.~R., {Bailes} M., {Dewey} R.~J.,  {Harrison} P.~A., 1993, MNRAS,
  263, 403

\bibitem[\protect\citeauthoryear{{Lorimer} et~al.}{{Lorimer}
  et~al.}{2006}]{lfl+06}
{Lorimer} D.~R. et~al., 2006, MNRAS, 372, 777

\bibitem[\protect\citeauthoryear{{Lorimer} \& {Kramer}}{{Lorimer} \&
  {Kramer}}{2005}]{lk05}
{Lorimer} D.~R.,  {Kramer} M., 2005, {Handbook of Pulsar Astronomy}.
\newblock Cambridge University Press

\bibitem[\protect\citeauthoryear{{Lorimer} et~al.}{{Lorimer}
  et~al.}{1995}]{lorimer1995}
{Lorimer} D.~R., {Yates} J.~A., {Lyne} A.~G.,  {Gould} D.~M., 1995, MNRAS, 273,
  411

\bibitem[\protect\citeauthoryear{{Lyne}}{{Lyne}}{1998}]{lyne1998}
{Lyne} A.~G., 1998, Advances in Space Research, 21, 149

\bibitem[\protect\citeauthoryear{Lyne, Manchester, \& Taylor}{Lyne
  et~al.}{1985}]{lmt85}
Lyne A.~G., Manchester R.~N.,  Taylor J.~H., 1985, MNRAS, 213, 613

\bibitem[\protect\citeauthoryear{Manchester et~al.}{Manchester
  et~al.}{2005}]{mhth05}
Manchester R.~N., Hobbs G.~B., Teoh A.,  Hobbs M., 2005, AJ, 129, 1993

\bibitem[\protect\citeauthoryear{Manchester et~al.}{Manchester
  et~al.}{2001}]{mlc+01}
Manchester R.~N. et~al., 2001, MNRAS, 328, 17

\bibitem[\protect\citeauthoryear{Manchester et~al.}{Manchester
  et~al.}{1996}]{mld+95}
Manchester R.~N. et~al., 1996, MNRAS, 279, 1235

\bibitem[\protect\citeauthoryear{{Maron} et~al.}{{Maron}
  et~al.}{2000}]{maron2000}
{Maron} O., {Kijak} J., {Kramer} M.,  {Wielebinski} R., 2000, A\&A, 147, 195

\bibitem[\protect\citeauthoryear{{Narayan}}{{Narayan}}{1987}]{narayan1987}
{Narayan} R., 1987, ApJ, 319, 162

\bibitem[\protect\citeauthoryear{{Perera} et~al.}{{Perera}
  et~al.}{2013}]{perera2013}
{Perera} B.~B.~P., {McLaughlin} M.~A., {Cordes} J.~M., {Kerr} M., {Burnett}
  T.~H.,  {Harding} A.~K., 2013, ApJ, 776, 61

\bibitem[\protect\citeauthoryear{{Possenti} et~al.}{{Possenti}
  et~al.}{2002}]{possenti2002}
{Possenti} A., {Cerutti} R., {Colpi} M.,  {Mereghetti} S., 2002, A\&A, 387, 993

\bibitem[\protect\citeauthoryear{Press et~al.}{Press et~al.}{1986}]{pftv86}
Press W.~H., Flannery B.~P., Teukolsky S.~A.,  Vetterling W.~T., 1986,
  Numerical Recipes: {T}he Art of Scientific Computing.
\newblock Cambridge University Press, Cambridge

\bibitem[\protect\citeauthoryear{Ridley \& {Lorimer}}{Ridley \&
  {Lorimer}}{2010}]{ridley2010}
Ridley J.~P.,  {Lorimer} D.~R., 2010, MNRAS, 404, 1081

\bibitem[\protect\citeauthoryear{{Schnitzeler}}{{Schnitzeler}}{2012}]{schnit2012}
{Schnitzeler} D.~H.~F.~M., 2012, MNRAS, 427, 664

\bibitem[\protect\citeauthoryear{{Smits} et~al.}{{Smits}
  et~al.}{2009a}]{smits2008}
{Smits} R., {Kramer} M., {Stappers} B., {Lorimer} D.~R., {Cordes} J.,
  {Faulkner} A., 2009a, A\&A, 493, 1161

\bibitem[\protect\citeauthoryear{{Smits} et~al.}{{Smits}
  et~al.}{2009b}]{smits2009}
{Smits} R., {Lorimer} D.~R., {Kramer} M., {Manchester} R., {Stappers} B., {Jin}
  C.~J., {Nan} R.~D.,  {Li} D., 2009b, A\&A, 505, 919

\bibitem[\protect\citeauthoryear{Taylor \& Manchester}{Taylor \&
  Manchester}{1977}]{tm77}
Taylor J.~H.,  Manchester R.~N., 1977, ApJ, 215, 885

\bibitem[\protect\citeauthoryear{{Weltevrede} \& {Johnston}}{{Weltevrede} \&
  {Johnston}}{2008}]{wj08}
{Weltevrede} P.,  {Johnston} S., 2008, MNRAS, 387, 1755

\bibitem[\protect\citeauthoryear{{Yusifov} \& {K{\"u}{\c c}{\"u}k}}{{Yusifov}
  \& {K{\"u}{\c c}{\"u}k}}{2004}]{yk2004}
{Yusifov} I.,  {K{\"u}{\c c}{\"u}k} I., 2004, A\&A, 422, 545

\end{thebibliography}
\bibliographystyle{mnras}

\label{lastpage}

\end{document}